\documentclass[manuscript]{aastex} 

\usepackage{graphics} 
\def\kms {$\rm km\,s^{-1}$}

\shorttitle{Stellar pop}
\shortauthors{A1 et al.}

\begin{document}


\title{2D mapping of young stars in the inner 180\,pc of NGC\,1068: correlation with molecular gas ring and stellar kinematics}


\author{Thaisa Storchi-Bergmann}
\affil{Universidade Federal do Rio Grande do Sul, IF, CP 15051, 91501-970, Porto
Alegre, RS, Brazil} \email{thaisa@ufrgs.br}

\author{Rogemar A. Riffel}
\affil{Universidade Federal de Santa Maria, Departamento de F\'\i sica, Centro de Ci\^encias Naturais e Exatas, 
97105-900, Santa Maria, RS, Brazil} 

\author{Rog\'erio Riffel}
\affil{Universidade Federal do Rio Grande do Sul, IF, CP 15051, 91501-970, Porto
Alegre, RS, Brazil}

\author{Marlon R. Diniz}
\affil{Universidade Federal de Santa Maria, Departamento de F\'\i sica, Centro de Ci\^encias Naturais e Exatas, 
97105-900, Santa Maria, RS, Brazil} 

\author{Tib\'erio Borges Vale}
\affil{Universidade Federal do Rio Grande do Sul, IF, CP 15051, 91501-970, Porto
Alegre, RS, Brazil}

\and

\author{Peter J. McGregor
} \affil{Research School of Astronomy and
Astrophysics, Australian National University, Cotter Road, Weston Creek,
ACT 2611, Australia}






\begin{abstract} 

We report the first two-dimensional mapping of the stellar population and non-stellar continua within the inner 180\,pc (radius) of NGC\,1068 at a spatial resolution of 8\,pc, using integral field spectroscopy in the near-infrared. We have applied the technique of spectral synthesis to data obtained with the instrument NIFS and the adaptive optics module ALTAIR at the Gemini North Telescope. Two episodes of recent star formation are found to dominate the stellar population contribution: the first occurred 300\,Myr ago, extending over most of the nuclear region; the second occurred just 30\,Myr ago, in a ring-like structure at $\approx$\,100\,pc from the nucleus, where it is coincident with an expanding ring of H$_2$ emission. Inside the ring, where a decrease in the stellar velocity dispersion is observed, the stellar population is dominated by the 300\,Myr age component. In the inner 35\,pc, the oldest age component  (age\,$\ge$\,2\,Gyr) dominates the mass, while the flux is dominated by black-body components with temperatures in the range 700\,$\le$\,T\,$\le$\,800\,K which we attribute to the dusty torus. We also find some contribution from black-body and power-law components beyond the nucleus which we attribute to dust emission and scattered light.

\end{abstract}


\keywords{galaxies: individual(NGC 1068) --- galaxies: active --- galaxies: nuclei -- galaxies: stellar population}



\section{Introduction}

NGC\,1068 is the closest, brightest and for this reason, the most studied Seyfert\,2 galaxy. Its distance of only 14.4\,Mpc \citep{crenkra00}  corresponds to a scale of 70\,pc\,arcsec$^{-1}$ at the galaxy. In spite of its proximity, there is only limited information about the stellar population in the vicinity of its active nucleus. 

\citet{lynds91} reported a high contrast structure surrounding the nucleus in a {\it Hubble Space Telescope} (HST) {\it Wide Field and Planetary Camera 2} (WFPC2) optical image of NGC\,1068 in the continuum (centered at 5470\,\AA) interpreted as a stellar cluster. \citet{thatte97a} reported the detection of this stellar cluster from stellar CO absorption features in the H and K band spectra and estimated its size as FWHM$\approx$\,50\,pc and age in the range $5-16\times10^8$\,yr. \citet{crenkra00} have analysed two long-slit spectra obtained with the  HST Imaging Spectrograph (STIS) covering the inner $\sim$500\,pc in the wavelength range 1150--10270\,\AA. They have found that the continuum in this inner region has contributions from two components: electron-scattered light from the hidden nucleus and stellar light. They conclude that the electron-scattered light dominates in the UV, while stellar light dominates at optical wavelengths. They found that the scattered light is strongest in a hot spot at $\approx$\,0\farcs3 northeast of the nucleus, being enhanced also in regions of strong narrow-line emission. The stellar component was found to be more concentrated southwest of the hot spot, approximately coincident with the nucleus, being dominated by an old (age\,$\ge\,10^9$\,yr) stellar population. 

\citet{crenkra00} also support the presence of a nuclear stellar cluster  $\sim$\,200\,pc in extent and argue that it cannot be too young (in the sense of having blue ionizing stars), because the light concentration seen in the optical continuum is not seen in the UV \citep{macchetto94}. The UV {\it Faint Object Camera} HST images of \citet{macchetto94} resemble instead the narrow-band [OIII] optical images, consistent with the interpretation that they are due to scattering, which is also supported by the fact that this light is highly polarized. As is well known by the AGN community, it was the discovery of this polarized light emission, showing a Seyfert\,1 spectrum, that led to the Unified Model of AGN \citep{antonucci85,antonucci93}. Another argument against the presence of ionizing stars is the absence, in the UV spectra of \citet{crenkra00}, of stellar absorption lines characteristic of the atmosphere of such stars.

More recently, \citet{emsellem06}, using optical observations with the SAURON  integral field spectrograph, have claimed that this galaxy is a candidate for a stellar velocity dispersion ($\sigma_\star$)  ``drop'' in the central $\sim$2 arcseconds: $\sigma_\star$ decreases to $\sim$\,100\,\kms\ in this region, being surrounded outwards by higher values in the range 150--200\,\kms.  This $\sigma_\star$-drop has been interpreted as the result of central gas accretion followed by an episode of star formation.

\citet{davies07} have also observed NGC\,1068 and other nearby AGN using the near-IR integral field spectrograph  SINFONI at the {\it Very Large Telescope} (VLT). In most of the galaxies they found signatures of circumnuclear starbursts with ages of 10--300\,Myr (i.e., no longer ionizing). In the case of NGC\,1068, they confirm the presence of a $\sigma_\star$-drop, and report also an excess of stellar continuum within the inner 1$^{\prime\prime}$, which they attribute to the presence of a nuclear disk that has experienced recent star formation. This would also explain the decrease in velocity dispersion.

\citet{martins10} performed stellar population synthesis on the
nuclear and extended regions of NGC\,1068 by means of near-infrared
spectroscopy ($0.8-2.4\rm \mu$m) using the NASA Infrared Telescope
Facility SpeX-XD data. A $0\farcs8\,\times\,15\arcsec$ slit oriented in the
north--south direction was employed. Their main result is that traces of young stellar population are found at $\sim$\,100\,pc south of the nucleus. They also report
a $\sim$\,25\% contribution of a power-law continuum at the nucleus and surrounding regions,
with excess contribution (up to 60\%) of this component at 100--150\,pc to both sides of the nucleus, associated by them with regions where new stars are being formed.  They also report that hot dust (1000\,K\,$\leq$\,T\,$\leq$\,1400\,K) has an important contribution to the nuclear
region while cold dust appears mostly to the south.  A
significant contribution of an intermediate-age stellar population to the
continuum, especially in the inner 200\,pc, was also found.

With the goal of mapping and quantifying the processes of feeding and feedback in the active nucleus of NGC\,1068, as well as its stellar population, we have analyzed two-dimensional spectroscopic data obtained with the Gemini North {\it Near-Infrared Integral Field Spectrograph} (NIFS)  in the J, H and K bands. In the present paper, we report the mapping and analysis of the stellar population and near-infrared continuum within the inner 180\,pc radius of NGC\,1068 at a spatial resolution of $\approx$\,8\,pc. We also include a brief analysis of the stellar kinematics, with the goal of relating it to the stellar population characteristics.

This paper is organized as follows: in Section 2 we discuss the observations and data reduction procedures, in Section 3 we derive the stellar kinematics, in Section 4 we obtain the properties the stellar population, in Section 5 we discuss the results, comparing the stellar population properties with the derived kinematics and the molecular gas distribution and in Section 6 we present our conclusions.

\section{Observations, data reduction and analysis}
\label{s:obs}

Spectra of NGC\,1068 were obtained with the Near-infrared Integral
Field Spectrograph (NIFS; McGregor et al. 2003) on the Gemini North
telescope in December 2006 under program GN-2006B-C-9. NIFS was used
with the ALTAIR facility adaptive-optics system in its natural guide
star mode. The compact Seyfert nucleus of the galaxy was used as the
adaptive-optics reference object. A standard star observed under the same
conditions as NGC\,1068 has a spatial profile with a FWHM of $0\farcs12$
at both $H$ and $K$, corresponding to a spatial resolution of $\sim$ 8
pc at the galaxy.

NIFS is an image-slicer integral-field spectrograph with a square
field of view of 3\arcsec$\times$3\arcsec, divided into 29 slitlets each
0$\farcs$103 wide with a sampling of 0$\farcs$044 along each slitlet. The
instrument was set to a position angle on the sky of 300\degr\ to
align the slitlets approximately with the axis of the radio jet and
ionization cone. This provided coarser spatial sampling along the jet
axis, and finer spatial sampling across the jet.

Observations at the $Ks$ grating setting were obtained on
12 December 2006 (UT). Similar observations at the $Kl$ grating
setting were obtained on 13 December 2006 (UT), in the $H$ band on 14
December 2006 (UT), and in the $J$ band on 15 December 2006 (UT). The
same orientation, observing procedure, exposure time, and standard
stars were used on each of these nights. The spectra used in the
present study cover the $H$ band from 1.48 \micron\ to 1.79
\micron\ and the $Kl$ band from 2.1 \micron\ to 2.5 \micron, both at a
spectral resolving power of $\sim$ 5290. The spectra in other
wavelength bands will be discussed in future papers.

Each data set consisted of multiple 90 s exposures. An offset sky
position was recorded first and was followed by nine galaxy fields obtained
on a $3\times3$ frame grid centered on the galaxy nucleus. The frame
offset was 1$\farcs$0, so frames were recorded at offsets of $-1\farcs0$,
$0\farcs0$, and $+1\farcs0$ with respect to the nucleus along and
perpendicular to the slitlets. This resulted in a full field of view (FOV)
$5\farcs0\times5\farcs0$ on the sky with maximum exposure in the
central $3\farcs0\times3\farcs0$ and less exposure in the corners of
the full field. The exposure grid was repeated nine times at
each grating setting for a total exposure time on any one of the nine
fields of 810\,s. No pixel-level dithering of the field positions was
employed. The nucleus of NGC\,1068 was within each field, and so was
subsequently used to spatially register each exposure. An argon/xenon
arc exposure was obtained immediately after the galaxy exposures were
completed. The A2 telluric standard star HIP\,5886 was observed
immediately prior to the NGC\,1068 observation, and the A0 star HIP\,18863 
was observed immediately after the NGC\,1068 observation. Both
were selected so that they were observed through approximately the same
atmospheric column as the NGC\,1068 observations. 

\begin{figure}
\centering
\includegraphics[scale=0.7]{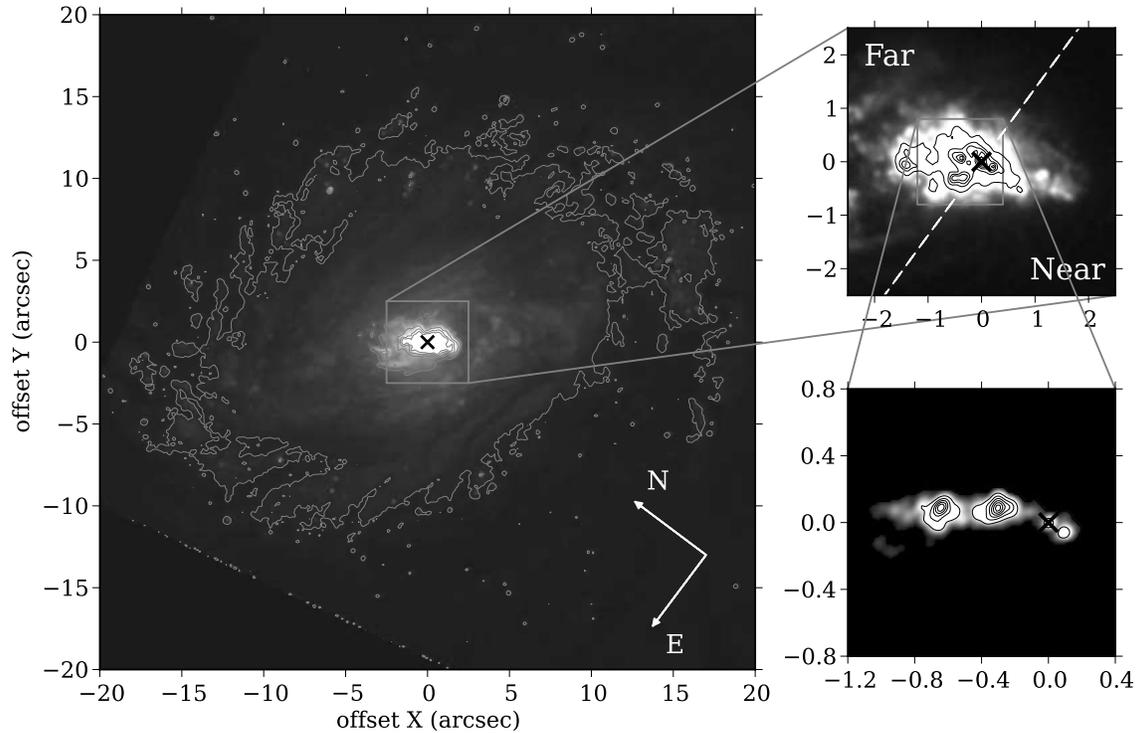}
\caption{The large left panel shows an HST image of NGC\,1068 through the filter F606W with the FOV of the NIFS observations indicated by the square. The top right panel shows an intensity map of the [O\,III]$\lambda$5007 emission line  from \citet{schmitt03}, as observed within the same FOV of the NIFS observations. The bottom right panel shows the radio jet. The black cross marks the position of  the galaxy nucleus.} 
\label{ima}
\end{figure}

The full FOV and orientation of our NIFS observations are illustrated by the white square over an {\it Hubble Space Telescope} (HST) image through the filter F606W in the left panel of Fig.\,\ref{ima}, where we also show, for reference, 
within our FOV, a narrow-band [O\,III]$\lambda$5007A image (top right panel of Fig.\,\ref{ima}) showing the known ionization cone \citep{schmitt03}, as well as the radio jet (bottom right panel).

The data reduction was performed using the GEMINI IRAF NIFS package
and followed standard procedures, as described in  \citet{sb09}. 
The nine sky exposures were first median-combined and dark
subtracted to form an average sky frame. Individual galaxy exposures
were then dark subtracted and the average sky exposure was
subtracted. The 29 two-dimensional slitlet spectra for each exposure
were cut from the raw data image and flatfielded. Known bad
pixels identified from the flatfield and dark exposures were
corrected by linear interpolation. Each two-dimensional slitlet spectrum
was then transformed to a rectilinear coordinate grid with horizontal
wavelength and vertical spatial axes using transformations derived from
the arc lamp and spatial calibration mask exposures. These transformed
spectra were then stacked in the second spatial dimension to form a
three-dimensional data cube with two spatial and one spectral axes.

The individual data cubes were then registered spatially by collapsing
each data cube in the spectral direction to form a continuum image,
determining the centroid of the NGC\,1068 nucleus in each image, and
then spatially shifting and median-combining all galaxy cubes using
the IRAF {\em imcombine} task. The resulting mosaiced data cube was
corrected for telluric absorption by multiplying each spectrum within
the cube by a one-dimensional correction spectrum derived from the
one-dimensional spectrum of the telluric standard star, HIP\,5886,
extracted from its data cube. A similar large-aperture one-dimensional
spectrum of the total light of HIP\,5886 was then used to flux
calibrate the mosaiced galaxy data cube. The intrinsic spectrum of the
standard star was set to the level of its 2MASS magnitude in the
wavelength band of the spectrum, and its shape was set to that of a
8600K blackbody, which has the same 2MASS $J-K$ color as HIP\,5886.

In order to have approximately square spaxels, we have then binned together two pixels together along the slices. The resulting calibrated data cube contains 2376 spectra
covering an angular region $4\farcs5\times5\farcs0$ (along the $x$ axis and $y$ axis, respectively) at a sampling of $0\farcs103\times0\farcs092$, which corresponds to $\sim 315\times350$
pc$^{2}$ (or a radius of $\sim$\,180\,pc about the nucleus) and $\sim
7.0\times6.4$ pc$^{2}$, respectively, at the galaxy.

\subsection{Data analysis}

\subsubsection{Line-of-Sight Velocity Distributions}

\begin{figure}
\centering
\includegraphics[scale=0.8]{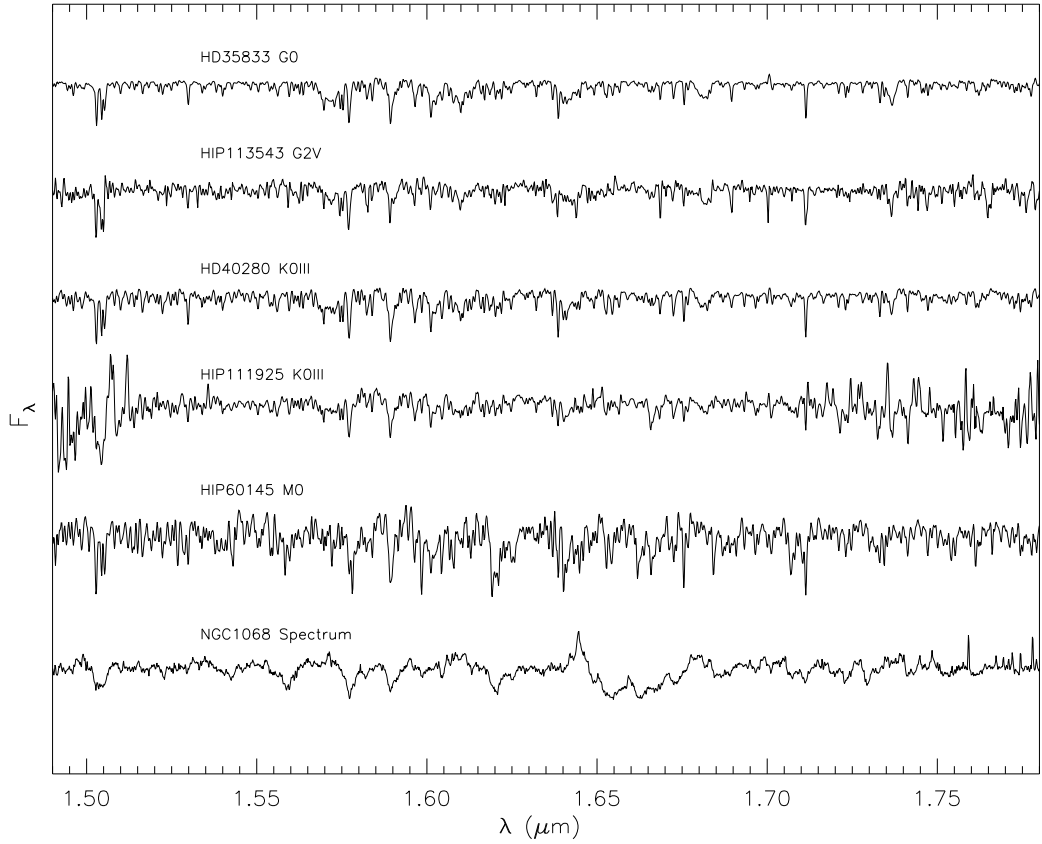}
\caption{H-band spectra of the stars used as templates by the pPXF, as well as sample spectrum of NGC\,1068 extracted at 0\farcs5 NE from the nucleus (position A from Fig.\,\ref{stel_fit}).} 
\label{template}
\end{figure}

In order to map the stellar kinematics, we have obtained the line-of-sight  velocity distributions (LOSVD) by fitting the absorption spectra in the H and K bands using the penalized Pixel-Fitting (pPXF) method of \citet{cappellari04}, following the procedure described in \citet{riffel08}. In summary, the pPXF method finds the best fit to a galaxy spectrum  by convolving stellar template spectra with a given LOSVD [$L(v)$], represented by Gauss-Hermite series \citep[e.g.][]{vandermarel93,gerhard93,profit}:
\begin{equation}
L(v)=\frac{{\rm e}^{-(1/2)/y^2}}{\sigma_*\sqrt{2\pi}}\left[1+\sum_{m=3}^M h_{m*} H_m(y) \right],
\end{equation}
where $y = (v -V_*)/\sigma$, $V_*$ is the stellar centroid radial velocity, $\sigma_*$ is the velocity dispersion, $v=c\,{\rm ln}\,\lambda$ and $c$ is the speed of light. The $H_m$ are the Hermite polynomials, $h_{m*}$ are the Gauss-Hermite moments \citep{cappellari04}. The pPXF routine outputs the stellar centroid velocity ($V_*$), velocity dispersion ($\sigma_*$) and the higher-order Gauss-Hermite moments $h_{3*}$ and $h_{4*}$. As discussed in \citet{riffel08} and  \citet{winge09}, the use of a library of stellar templates, instead of a single stellar spectrum, is fundamental for a reliable derivation of the stellar kinematics. In the K-band, we have used as templates the spectra of 60 stars from the the  Gemini library of late spectral type stars observed with the Gemini Near-Infrared Spectrograph (GNIRS) IFU and  NIFS \citep{winge09}. In the H-band, we do not have a large number of templates available, and we have thus used the 
spectra of five late type stars with public NIFS data in the Gemini archive. The H-band spectra of these stars are shown in Fig.\,\ref{template}, together with the NGC\,1068 nuclear spectrum.

\subsubsection{Spectral Synthesis}

In order to map the distribution of the continuum and stellar population properties, we have  used the {\sc starlight} spectral synthesis code \citep{asari07,cid05,cid05a} adapted to spectra in  the near-IR \citep{rogerio09}. This code fits a galaxy's spectrum  by a combination of  elements from a  ``spectral base" (described below),  giving as output the weights of each base element in the synthetic spectrum \citep{cid09,rogerio09}.

The spectral base was constructed with the most recent Evolutionary Population Synthesis  models of \citet[][see also \citet{maraston98,maraston05}]{maraston11} following \citet{rogerio09}, and comprise single stellar populations (SSPs) synthetic spectra covering 12 ages ($t=$\,0.01, 0.03, 0.05, 0.1, 0.3, 0.5, 0.7, 1, 2, 5, 9 and 13\,Gyr) and solar metallicity.  We did not include a stellar population younger than 10\,Myr as we verified it was not necessary because it becomes degenerated with a power-law component \citep{sb00} also included in the synthesis. The power-law ($F_\nu\propto\nu^{-1.5}$) was used to represent  the possible contribution of a featureless continuum from the active nucleus \citep[e.g.][]{cid04}. We also included black-body functions for temperatures in the range 700-1400\,K  (one for each 100\,K interval) in order to account for possible contributions from dust emission.

Following \citet{cid04}, we have binned the contribution of the SSP $x_j$ elements into stellar population components of four age ranges $t_j$:  
young ($x_y$: $t_y\leq 50$~Myr); young-intermediate ($x_{yi}$: $0.1 \leq t_{yi} \leq 0.7$~Gyr), intermediate-old  ($x_{io}$: $1 \leq t_{io} \leq 2$~Gyr)  and old  ($x_{o}$: $5 \leq t_o \leq 13$~Gyr). We present the synthesis results in two forms: percent flux contribution of each SPC at $\lambda=\rm 2.1 \mu m$ (x$_j$) and percent mass contribution to the total mass (m$_j$).

\section{Results: Stellar Kinematics}

As described in the previous section, we have used the penalized Pixel-Fitting (pPXF) method of \citet{cappellari04}  to fit the absorption spectra in the H and K bands in order to obtain the line-of-sight  velocity distributions (LOSVD). In our previous studies \citep[e.g.][]{riffel08,riffel10, riffel11} we have used only the K-band spectra for the derivation of the stellar kinematics. But in the case of NGC\,1068, the K-band spectra within $\approx$\,0\farcs3 from the nucleus are strongly diluted by a red continuum that can be attributed to emission by hot dust from the circum-nuclear torus, as in the case of NGC\,4151 \citep{riffel09}. In addition, the stellar K-band absorption features (CO bands)  towards the north-west, where the ionization cone is located \citep{crenkra00}, are contaminated by strong emission from the  [Ca\,{\sc viii}]$\lambda2.322\,\mu$m coronal line. In order to avoid these contaminations present in the K-band spectra, we have used the H-band spectra to derive the stellar kinematics for angular distances from the nucleus smaller than 0\farcs3. For the other regions we used the K-band spectra whenever the contamination by the  [Ca\,{\sc viii}]$\lambda2.322\,\mu$m coronal line was negligible, as the stellar absorption features are stronger in the K-band than in the H-band, resulting in a better constrained stellar population, particularly for the outermost regions. In order to ``automatically'' select which band to use for the derivation of the stellar kinematics, we used the flux of the [Ca\,{\sc viii}]$\lambda2.322\,\mu$m coronal line: in regions with fluxes of this line higher than $3\sigma$ ($\sigma$ here meaning the root mean square deviation of the continuum flux in the neighboring continuum), we used the H-band spectra, while in other regions we used the K-band spectra. 

\begin{figure*}
\vspace{-0.5cm}
\begin{minipage}{1\linewidth}
\hspace{0.5cm}
\includegraphics[scale=0.37]{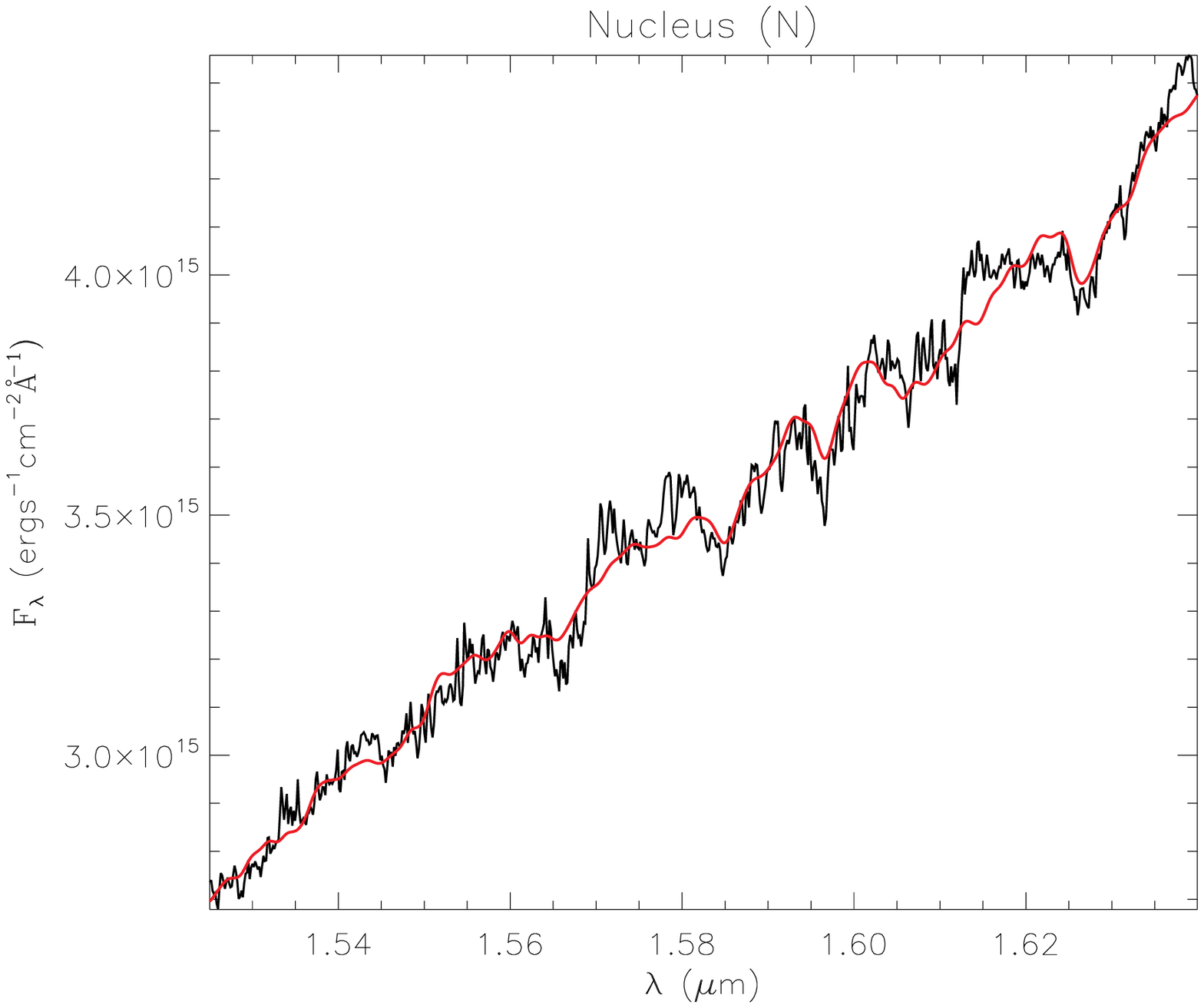}
\includegraphics[scale=0.37]{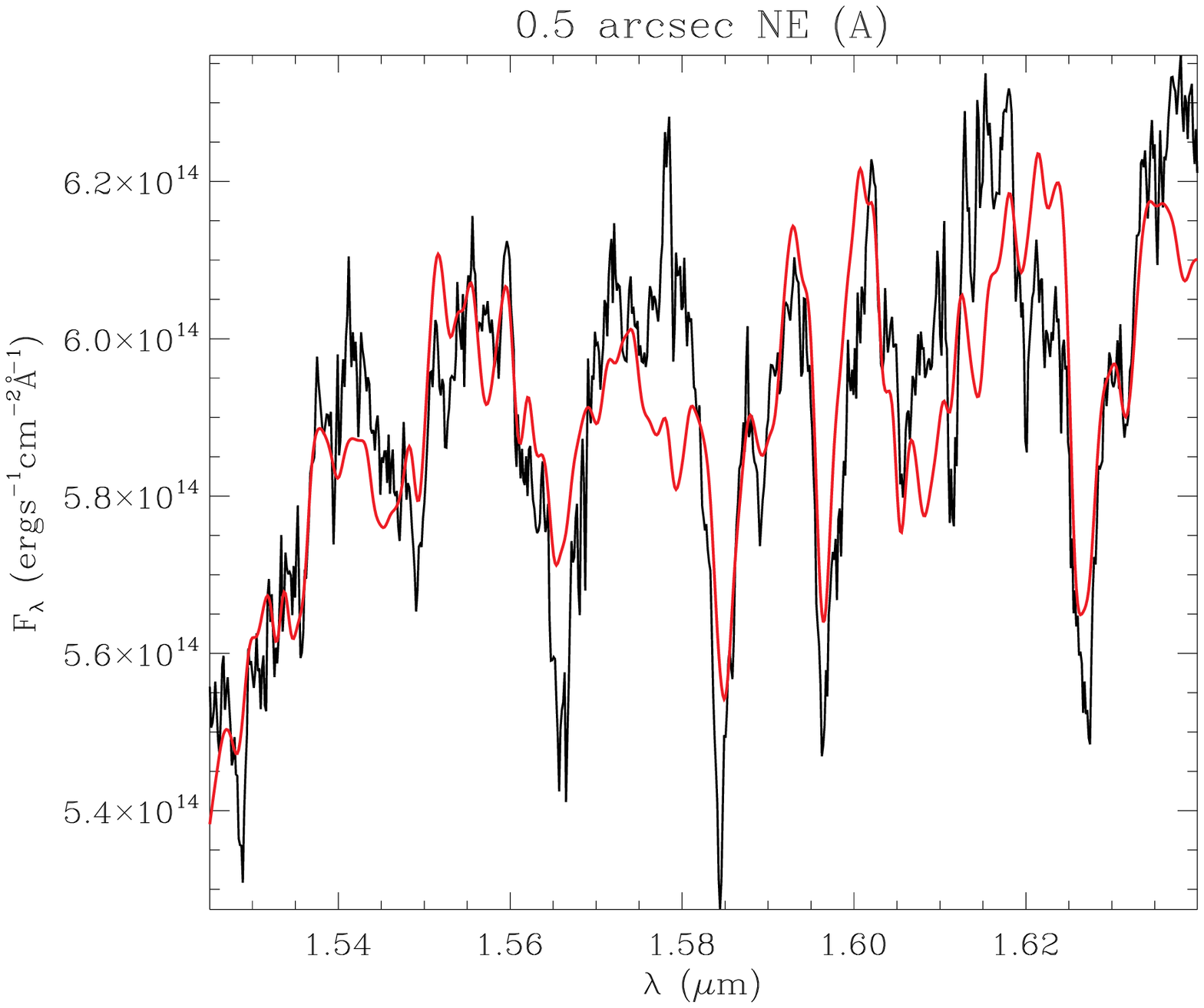}
\end{minipage}
\vspace{-0.5cm}
\hspace{-0.85cm}
\includegraphics[scale=0.8]{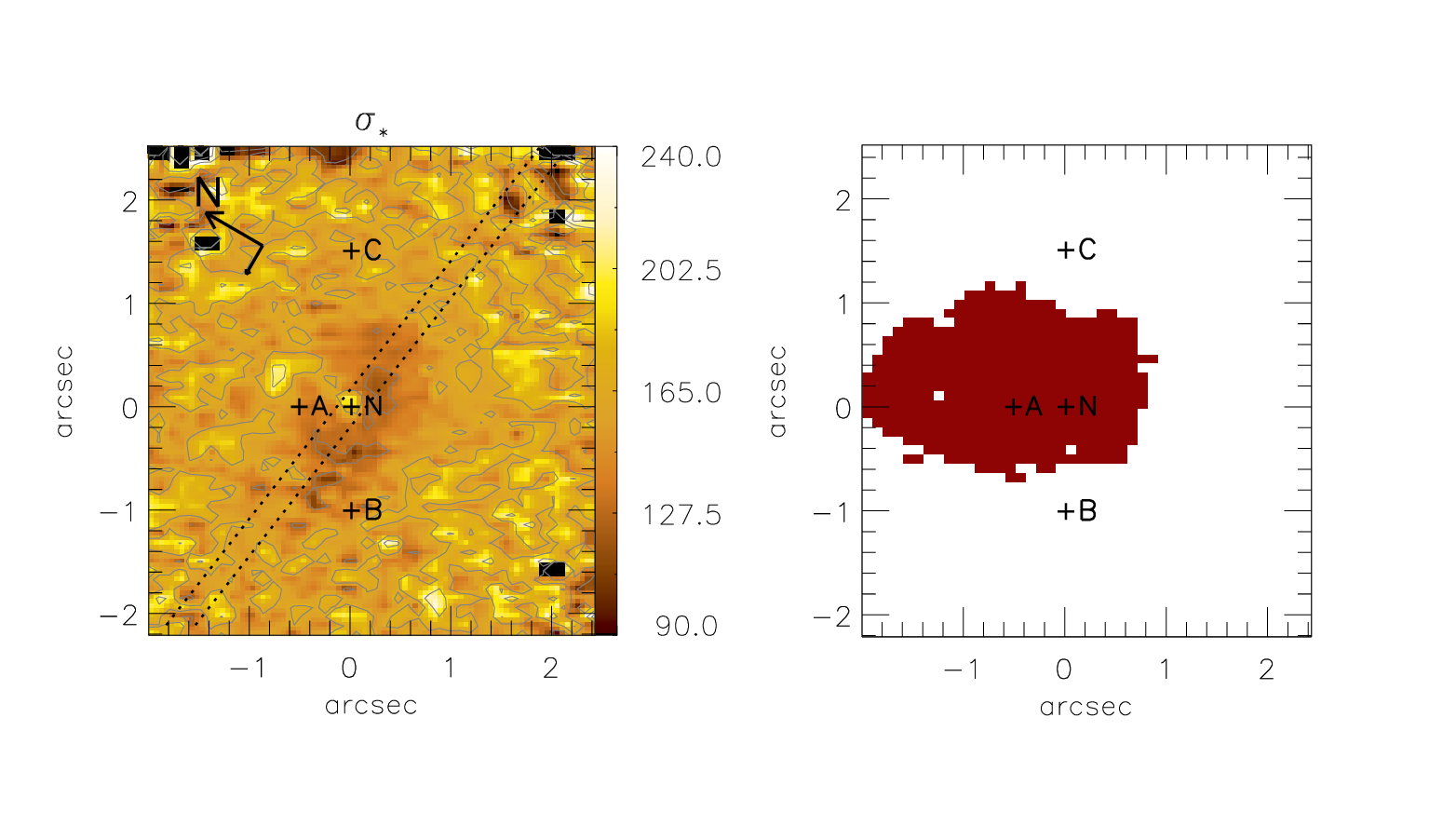}
\begin{minipage}{1\linewidth}
\vspace{-0.5cm}
\hspace{0.5cm}
\includegraphics[scale=0.37]{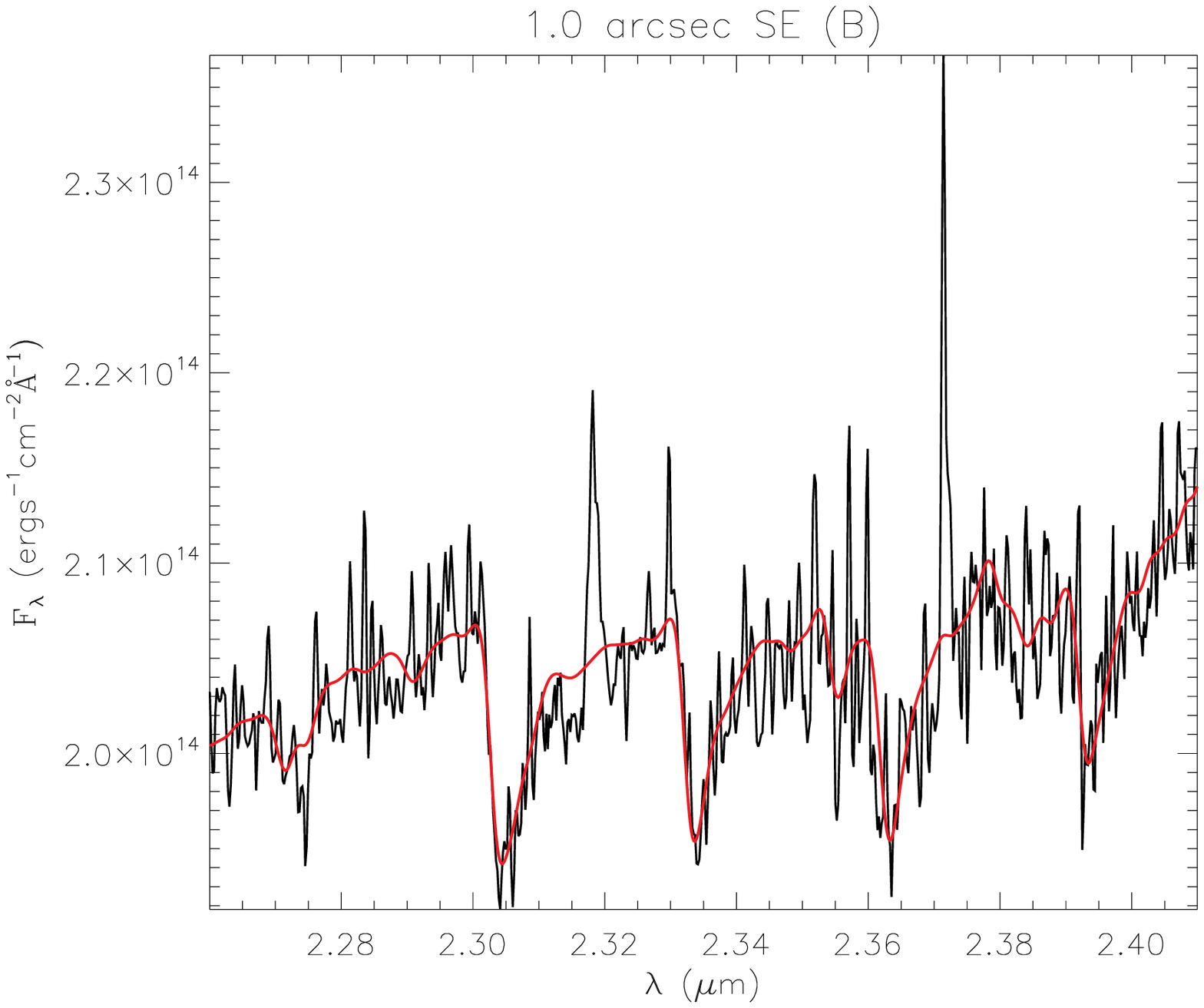}
\includegraphics[scale=0.37]{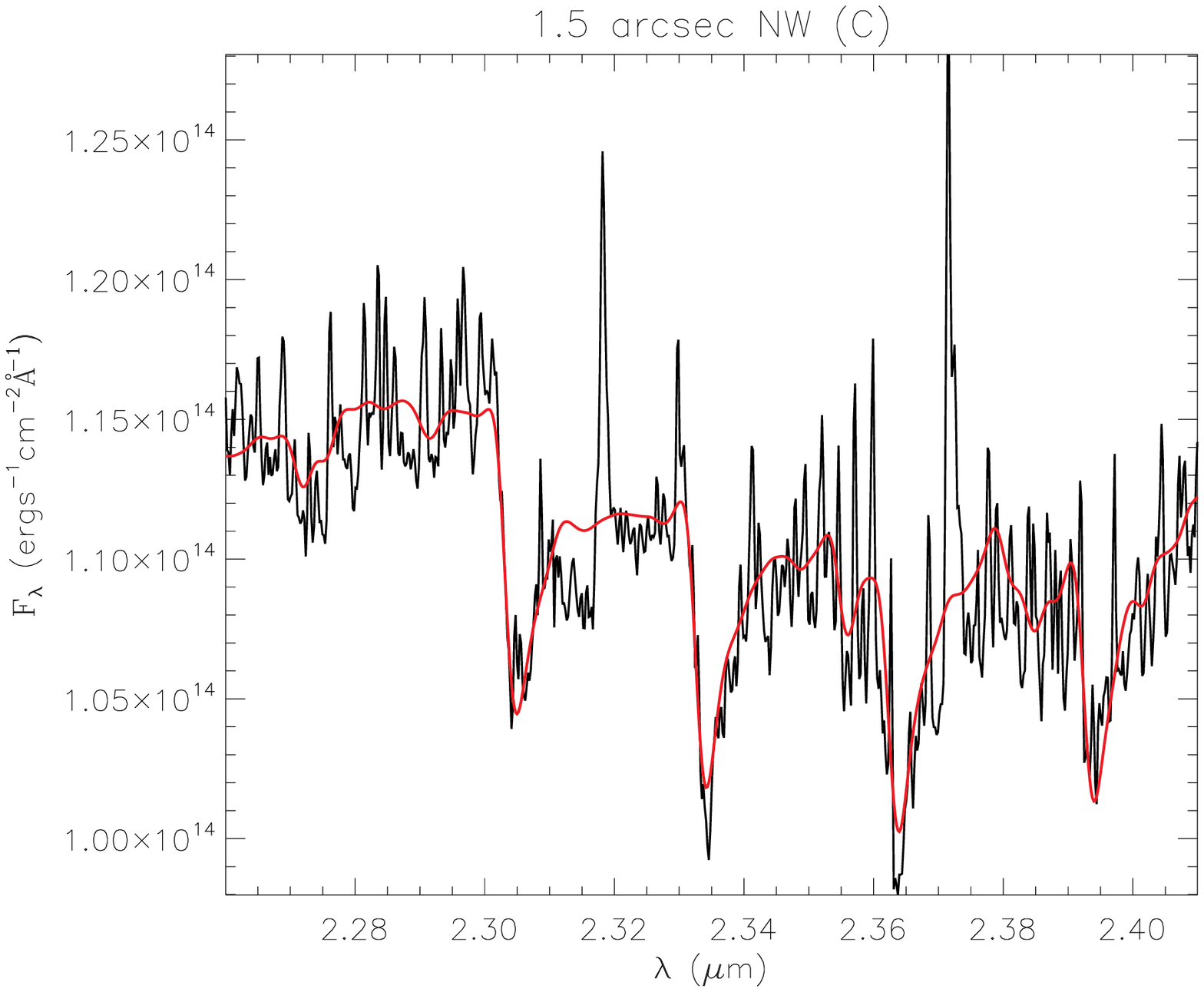}
\end{minipage}
\caption{Top and bottom: sample spectra showing the fits obtained via the Penalized Pixel Fitting method (in red) compared to the observed spectra (in black) at locations identified in the central panels. Center left: velocity-dispersion map showing the pseudo-slit along the major axis. Center right: ``mask" showing the region where the kinematics were derived from the H-band spectra.}
\label{stel_fit}  
\end{figure*}

In the top and bottom panels of Figure\,\ref{stel_fit}, we show representative fits (in red) compared to the observed spectra (in black), extracted at locations identified in the resulting velocity dispersion map: the nucleus, 0\farcs5\,north-west,  1\farcs0\,south-east and 1\farcs5\,north-west of the nucleus.  The region where we have used the H-band spectra for the derivation of the kinematics is shown in the center right panel of Fig.\,\ref{stel_fit} (shaded region in the figure).

\begin{figure}
\centering
\vspace{-1.6cm}
\includegraphics[scale=0.65]{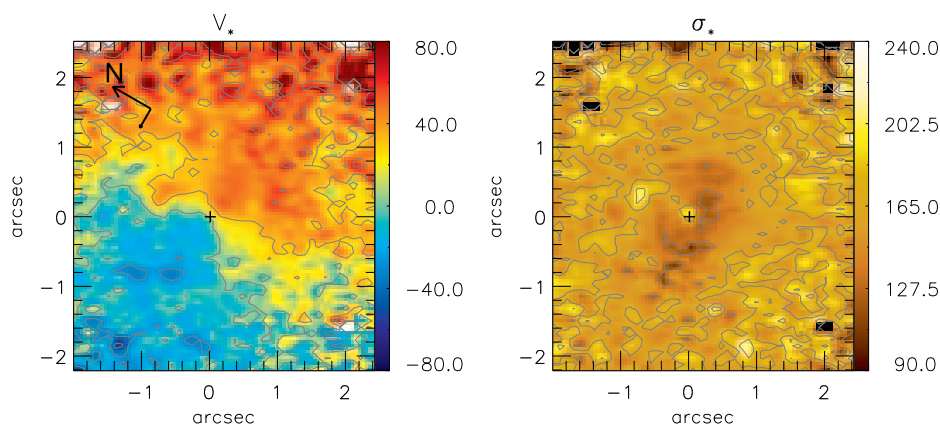}
\caption{Stellar kinematics obtained from the fit of the stellar population features in the H and K bands (see Fig.\,\ref{stel_fit}). Left: centroid velocity map; right: velocity dispersion map. The central cross marks the position of the nucleus and the color bars show velocity values in km\,s$^{-1}$. The orientation is indicated by arrows in the left frame.} 
\label{stel}
\end{figure}

\begin{figure*}
\vspace{-0.5cm}
\begin{minipage}{1\linewidth}
\hspace{1cm}
\includegraphics[scale=0.65]{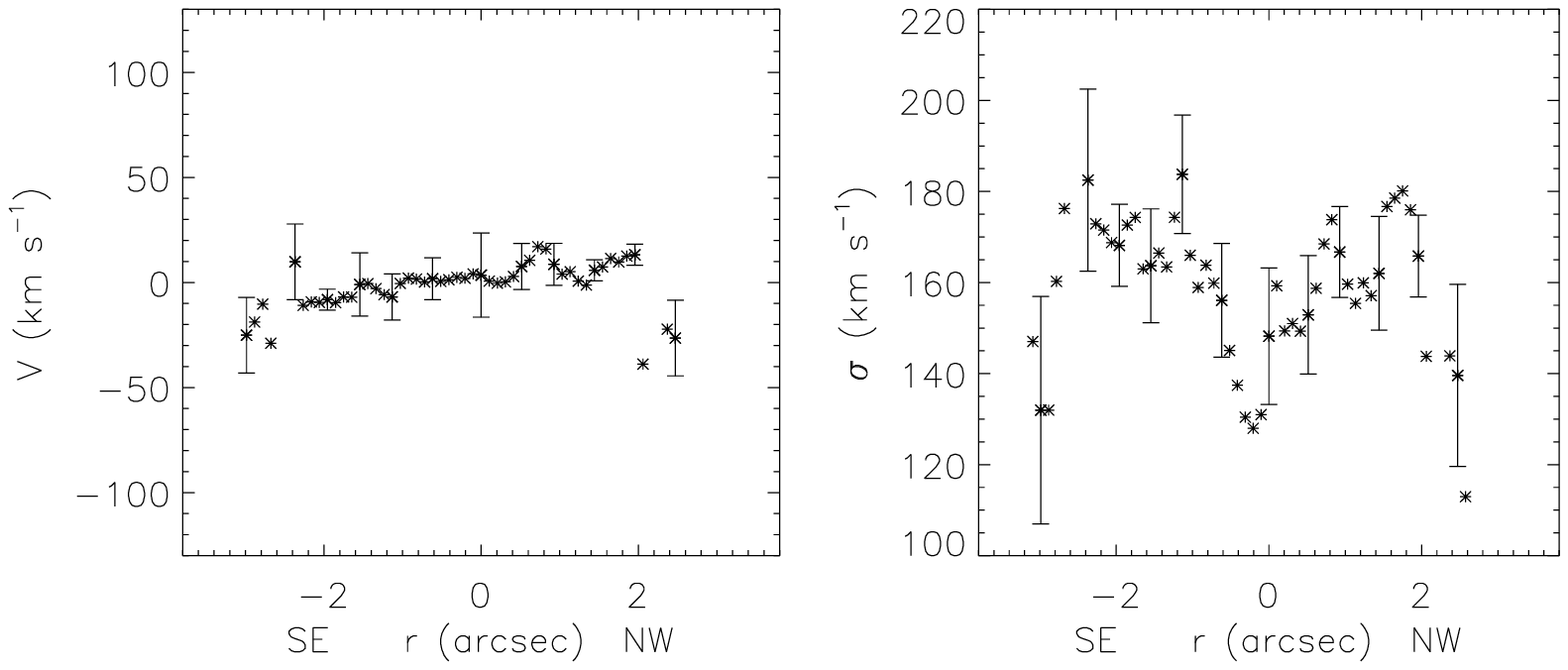}
\end{minipage}
\begin{minipage}{1\linewidth}
\hspace{1cm}
\includegraphics[scale=0.65]{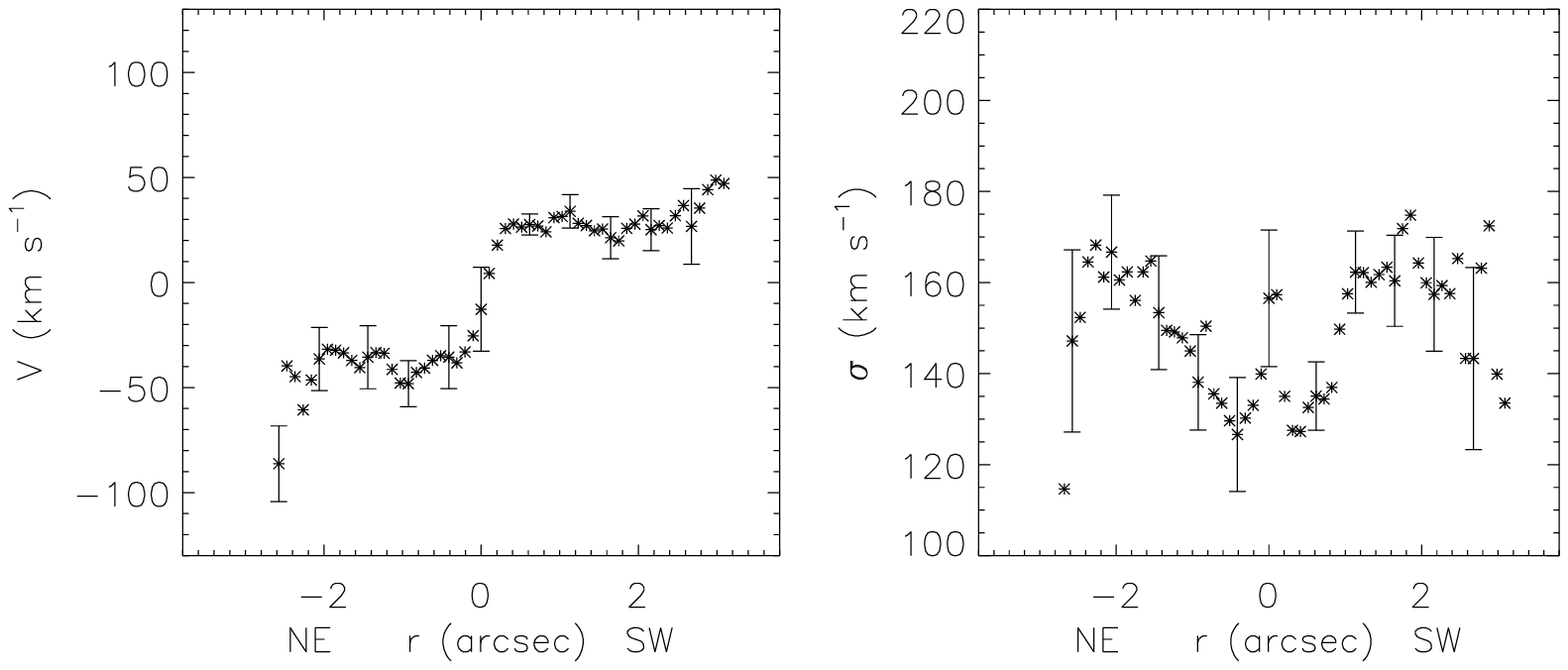}
\end{minipage}
\caption{Stellar kinematics extracted along two pseudo-slits shown together with typical error bars. Left: centroid velocity; right: velocity dispersion. Top: along the galaxy minor axis (PA$=-10^\circ$); bottom: along the major axis (PA=80$^\circ$).} 
\label{cut}
\end{figure*}

The  left panel of Fig.\,\ref{stel} shows the centroid velocity map, from which we subtracted the systemic velocity of $V_{\rm LSR}=1168\pm15\,{\rm km\,s^{-1}}$ [corrected for the motion relative to the local standard of rest (LSR)] adopted as the average stellar velocity within an aperture of 0$\farcs$5\,$\times$\,0$\farcs$46 centered on the nucleus. The centroid velocity map shows a rotation pattern, with the kinematic major axis approximately along the east--west direction, at PA$\approx\,80^\circ$, with blueshifts to the east and redshifts to the west, reaching a projected velocity amplitude of 45\,\kms. One-dimensional cuts showing the centroid velocities along  two ``pseudo-slits'' with width 0$\farcs$3 oriented along the galaxy major and minor axes (PA$\approx\,80^\circ$ and $\approx\,-10^\circ$, respectively) are shown in Fig.\,\ref{cut}. The pseudo slit along the major axis is illustrated in the central left panel of Fig.\,\ref{stel_fit} by the dashed lines.

The right panel of Fig.\,\ref{stel} shows the velocity dispersion $\sigma_\star$  map. The highest values reach  $\sigma_\star\,\approx\,190\,{\rm km\,s^{-1}}$, observed right at the nucleus and in patchy regions farther than 1$\arcsec$ from the nucleus towards the borders of the field. Within a region between $\approx$\,0$\farcs2$ and $\approx$\,1\arcsec\ from the nucleus (corresponding to  14\,pc  and 70\,pc, respectively, although more extended along east--west than along north-south), $\sigma_\star$ decreases to 125\,\kms. One-dimensional cuts showing the velocity dispersions along the  two ``pseudo-slits'' -- passing through the galaxy major and minor axes are shown in Fig.\,\ref{cut}.

In order to calculate the errors in the centroid velocity and velocity dispersions, we have performed Monte Carlo simulations in which Gaussian noise was added to the spectra. The corresponding error bars are illustrated in the one dimensional cuts of Fig.\,\ref{cut}, showing that the errors both in the centroid velocities and in the velocity dispersions range from 10 to 20 km\,s$^{-1}$.

\section{Results: Stellar population}

As described in Sec.\,\ref{s:obs}, we have used the {\sc starlight} spectral synthesis code adapted to spectra in  the near-IR \citep{rogerio09} to fit the galaxy's spectra  by a combination of  templates from a base including  black-body and power-law continua and single age stellar populations,  giving as output the contribution of each base element to the synthetic spectrum \citep{cid09,rogerio09}.

\begin{figure}
\centering
\includegraphics[scale=0.6]{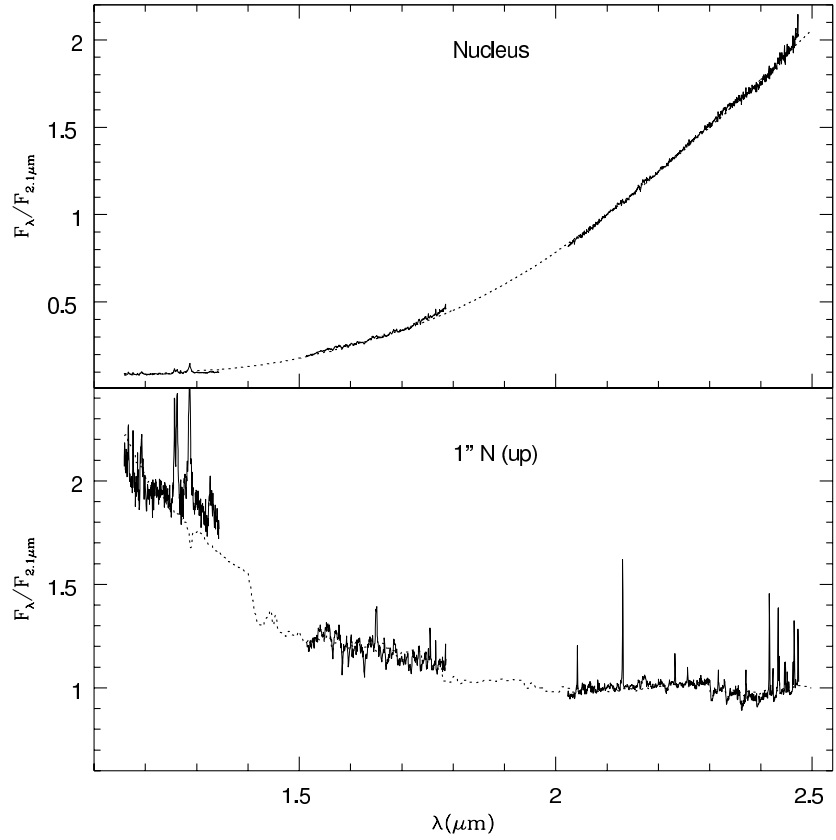}
\caption{Fit of the combined stellar population and continuum components to the nuclear spectrum (top) and to an extra-nuclear one at 1 \arcsec (70\,pc) north of the nucleus  (bottom). The observed spectra are shown as continuous lines and the fits as dashed lines. The nuclear spectrum is dominated by the contribution of the blackbody functions corresponding to the dust emission.} 
\label{fitting}
\end{figure}

We show in Fig.\,\ref{fitting}, the results of the synthesis for two spectra (extracted within our aperture of 0\farcs1$\times$0\farcs09), one  at the nucleus (top panel) and the other at 1$^{\prime\prime}$ north of it (bottom panel). The observed (continuous line) and synthetic (dashed line) spectra were normalized to unity at 2.1\,$\mu$m. 


\begin{figure}
\centering
\includegraphics[scale=0.5]{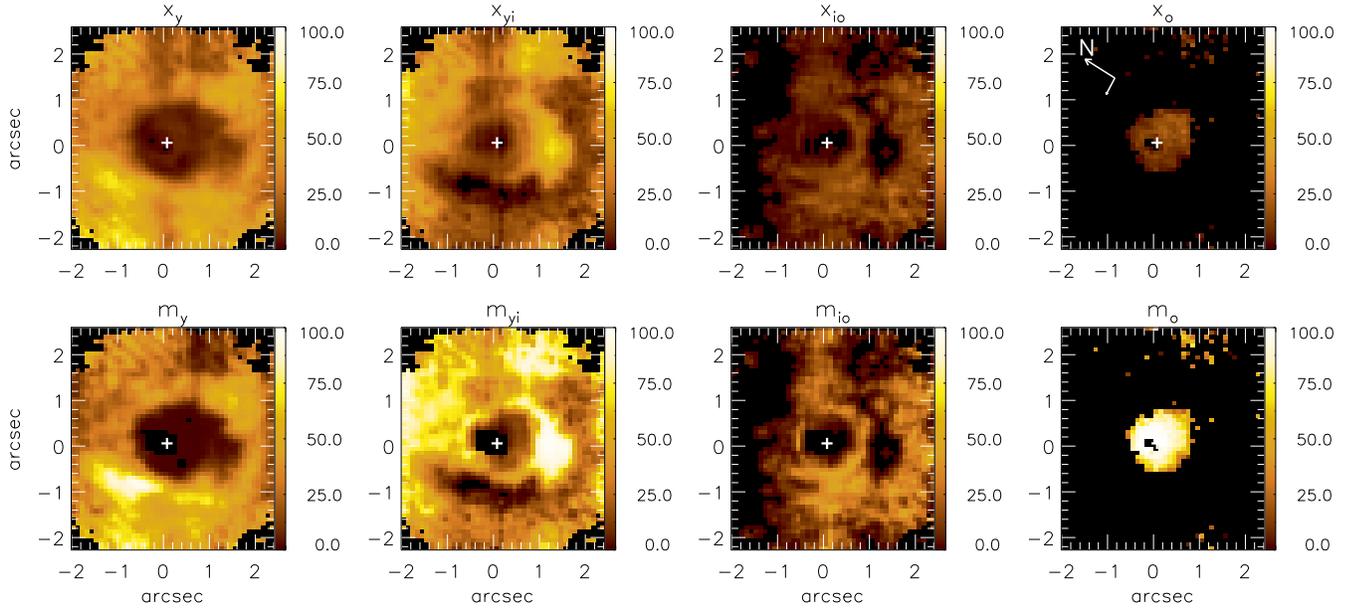}
\caption{Top: spatial distributions of the percentage contribution of each SPC to the flux at $\lambda=2.1\,\mu$m.  From left to right: young ($\leq$\,100\,Myr), young-to-intermediate (0.3--0.7~Gyr), intermediate-to-old (1--2~Gyr) and old (5--13~Gyr). Bottom: spatial distributions of the mass-weighted contribution from each SPC.  } 
\label{pop}
\end{figure}

The spatial distribution of the stellar population components are shown in Fig.\,\ref{pop}. The top panels show the percent contributions to the light at $\lambda$2.1$\mu$m of the young, young-intermediate, intermediate-old and old age components, from left to right:  x$_y$, x$_{yi}$, x$_{io}$ and x$_o$, while the bottom panels show the corresponding mass contributions m$_y$, m$_{yi}$, m$_{io}$ and m$_o$. We have masked out  from the figures the results of the synthesis in the regions for which \textit{adev} (see below) is smaller than 2\% in order to show only the results based on robust fits.


The flux contribution to the continuum within the inner 180\,pc of NGC\,1068 is dominated by the young and intermediate age stellar populations.The flux contribution of the young population x$_y$, as well as its contribution to the total mass m$_y$, shows a ``hole'' of radius $\approx$\,1\arcsec\ around the nucleus covering the  east--north--west directions, but extending almost to $\approx$2\arcsec\ towards the soutwest--south. The x$_y$ and m$_y$ are distributed around this ``hole", with enhanced contributions in a partial ring around the nucleus from the northeast-east towards the south and to the west-southwest. The young-intermediate age contribution x$_{yi}$ and m$_{yi} $ shows also a ``hole" around the nucleus but with approximately half the radius of the one observed in x$_y$ and m$_y$, and with similar extent in all directions. The southwest--south extension of the ``hole'' seen in x$_y$ and m$_y$ is not present in x$_{yi}$ and m$_{yi}$, which show enhanced contributions in this region -- a clump between $\approx$\,1 and 2\arcsec\ to the right of the nucleus in Fig.\,\ref{pop} -- where m$_{yi}\,\approx$\,100\%. The remaining regions of enhanced contributions of x$_{yi}$ and m$_{yi}$ seem to be anti-correlated with those of enhanced x$_y$ and m$_y$,  being observed mostly to the north-east through the north and north-west. 

The contribution  of the old stellar population, x$_o$ and m$_o$, is restricted to within $\approx$0$\farcs$5 (35\,pc) from the nucleus, where the contribution to the total mass m$_o$ is dominated by this component. The contributions of the intermediate-old stellar population x$_{io}$ and m$_{io}$ are small and avoid the nuclear region, being somewhat co-spatial with the regions of enhanced contribution of the young stellar population.

\begin{figure}
\centering
\includegraphics[scale=0.5]{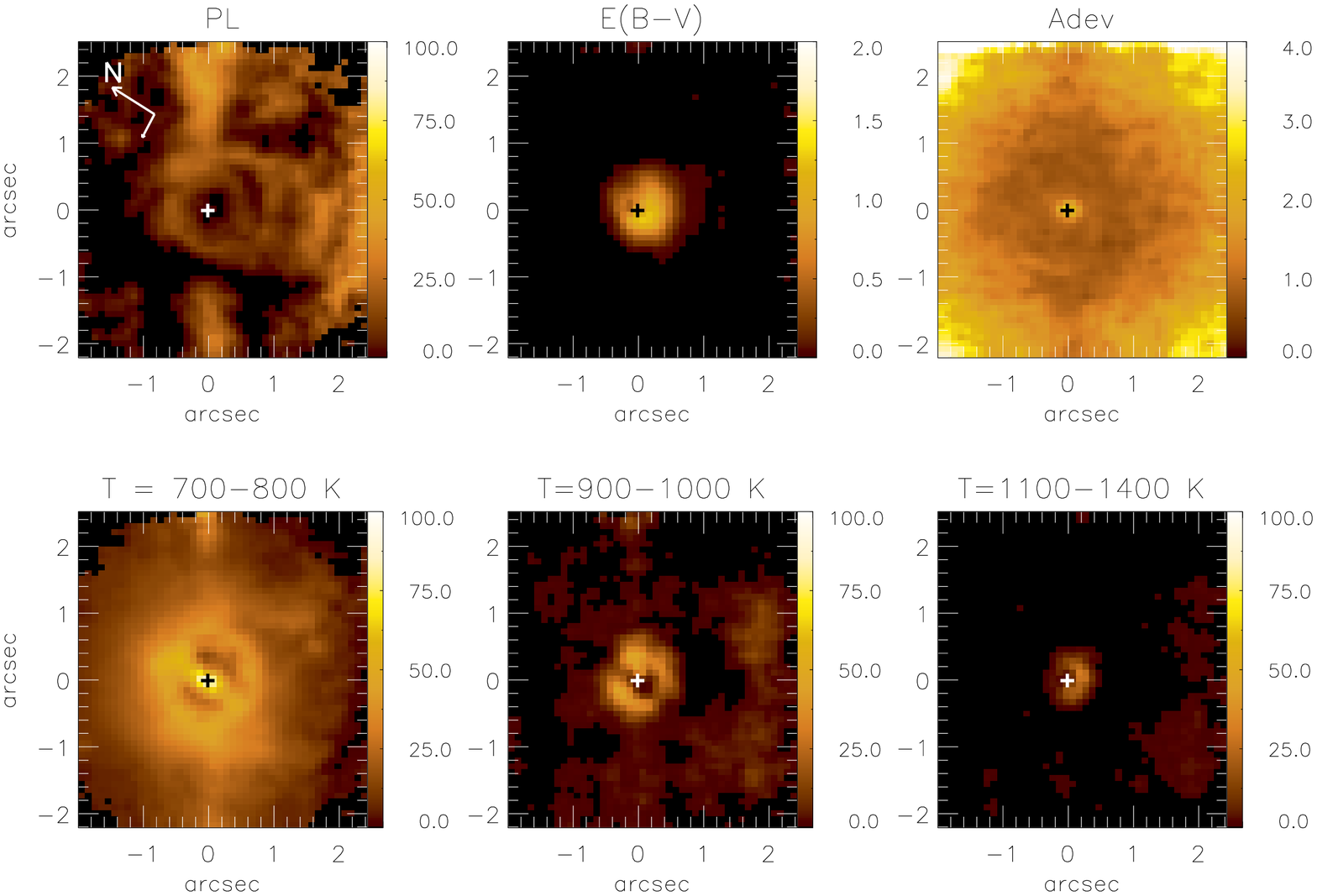}
\caption{Top, from left to right, distributions of: (1) percent flux contributions of the power-law component;  (2)reddening E(B-V) derived from the stellar population synthesis;  (3) the parameter Adev, which quantifies the quality of the spectral fit. Bottom, from left to right: contributions of the blackbody components of temperature  ranges shown at the top of each panel.}
\label{pop-ebv}
\end{figure}

Fig.\,\ref{pop-ebv} shows the contribution of the power-law and black-body components, together with the reddening derived from the stellar population synthesis. The rightmost panel shows the distribution of the \textit{adev}, which is the percent mean deviation of the fit -- $|O_\lambda-M_\lambda|/O_\lambda\times\,100$,  where $O_\lambda$ is the observed spectrum and $M_\lambda$ is the fitted model \citep{cid04,cid05} -- and gives the robustness of the fit. Values of $adev\lesssim\,2$\% correspond to good fits, while higher values, observed towards the corners of the field, indicate less robust fits.

Fig.\,\ref{pop-ebv} shows that  the contribution to the flux within 0\farcs5 of the nucleus is dominated by the black-body components with 700\,$\le$\,T\,$\le$\,800\,K, which also show a significant contribution to a radius $\approx$\,1\arcsec.The black-body components with 900\,$\le$\,T\,$\le$\,1000\,K also have some contribution closer to the nucleus, while for T\,$\ge$\,1100K the contribution occurs even closer to the nucleus but is of negligible strength.
Fig.\,\ref{pop-ebv} also shows that the reddening obtained from the synthesis is only significant within the inner $\approx$ 0$\farcs$5, where it reaches E(B-V)=1.5\,mag. The power-law component is extended towards the south-southeast (right in Fig.\,\ref{pop-ebv}), while the vertical structure is an artifact probably due to scattering of the strong nuclear source along the NIFS mirror slices.

\section{Discussion} \label{discussion}

\subsection{Stellar kinematics}

Fig.\,\ref{stel} shows a strong signature of rotation about the nucleus and Fig.\,\ref{cut} reveals a steep rotation curve, with the line-of-sight velocity reaching $\approx$\,40\,km\,s$^{-1}$ at just 0\farcs2, or 140\,pc from the nucleus. 
This result can be compared with those of previous studies. The first two-dimensional study of the stellar kinematics of the nuclear region of NGC\,1068  was done by \citet{garcia-lorenzo97} using the instrument 2D-FIS on the {\it William Herschell Telescope} (WHT). 95 fibers, each of 0\farcs9 diameter, were used to cover the inner 9$\farcs2\times12\farcs$2 at 5\AA\  spectral resolution in order to map the stellar kinematics in the Ca\,II triplet around 8500\AA. The main result of this paper was that the stellar kinematics of the inner 3\arcsec\ has a center which does not agree with the center of the stellar kinematics at radii larger than  5\arcsec, suggesting a decoupled kinematics between the the inner and outer regions. Our observations are restricted only to the inner region, thus we cannot verify the claimed decoupling. But in the inner region, the velocity amplitude of  $\approx$\,50\,km\,s$^{-1}$, and the orientation of the major axis, of 88$^\circ\pm5^\circ$, are in agreement with our velocity field shown in Fig.\,\ref{stel}. Their velocity dispersions -- in the range 250--300\,\kms -- are somewhat larger than our values (120--180\kms). This difference is probably due to the different methods and stellar template spectra used to obtain the velocity dispersion values. 

We can also compare our results with those from  \citet{davies07}, who observed the inner $\approx$\,2\arcsec\ of NGC\,1068 at 0\farcs1 resolution in the H-band with SINFONI at the VLT. They show in their Fig.\,21 radial profiles of the stellar velocity and velocity dispersion. The radial profile of the stellar velocity curve is very steep showing an increase in the stellar velocity  from zero to 40--50\,km\,s$^{-1}$ at 0\farcs3 of the nucleus, which is in good agreement with our velocity curve along P.A.$=80^\circ$ (although they do not show the two-dimensional velocity field). \citet{davies07} also report a decrease of $\sigma_\star$ from 130\,km\,s$^{-1}$ at 1\arcsec--2\arcsec\  to only 70\,km\,s$^{-1}$ at the very center. Our $\sigma_\star$ measurements shown in Figs.\,\ref{stel} and \ref{cut} support the presence of a decrease in the velocity dispersion within the inner 2\arcsec. But our values of $\approx$\,165\,km\,s$^{-1}$ at 2\arcsec\ down to $\approx$\,125\,km\,s$^{-1}$ at 0\farcs5 are larger than those reported by \citet{davies07}. Again we attribute this difference to different methods to obtain the velocity dispersion values as well as to the distinct stellar templates used by \citet{davies07} when compared to ours. 

\citet{davies07} have interpreted their observations of the rotation pattern plus low $\sigma_\star$ within the inner $\approx$\,1\arcsec\ as being due to a compact stellar disk with $\approx$\,70\,pc radius that has maintained the ``cold" kinematics of the gas from which these stars have formed. They also claim  that the stellar continuum increases by a factor of 2 above the inward extrapolation of the galaxy profile and that these two signatures are evidences of the presence of a nuclear disk which has experienced recent star formation. They have derived an inclination of 40$^\circ$ and a position angle of P.A.$=85^\circ$ for the compact disk. This position angle is close to the major axis of the galaxy and the orientation and extent of the disk are consistent with the orientation and extent of the distribution of low $\sigma_\star$ values shown in our Fig.\,\ref{stel}.

Another comparison we can make is with the results reported by \citet{emsellem06}, who have mapped the stellar velocity field in the inner $\approx$\,20\arcsec of NGC\,1068 using optical observations with SAURON, although with lower spatial resolution (apertures of $\approx$\,3\arcsec). They report a major axis position angle of 80$^\circ$, derived from the stellar kinematics and which is also in agreement with the photometric major axis of the outer galaxy disk. This PA is consistent with our kinematics shown in Fig.\,\ref{stel}.
Their stellar velocity dispersion values are also similar to ours and they claim the presence of a $\sigma-$drop  within the inner 3\arcsec, from values of $\approx$150\,km\,s$^{-1}$ down to $\approx\,100$\,km\,s$^{-1}$. This is also in approximate agreement with our findings, only that we resolve better the ``$\sigma-$drop'', which we find to be within the inner arcsecond (70\,pc).


\subsection{Stellar population}

\subsubsection{Young stars and correlation with H$_2$ ring}

Fig.\,\ref{pop} shows that the main stellar population  components contributing to the flux within the inner 180\,pc  of NGC\,1068 are the young and the young-intermediate ones. Within these components, we can verify from the output of the synthesis that the age bins which contribute with most of the flux are those of 30\,Myr  and 300\,Myr, respectively. Thus, it can be concluded that two main episodes of recent star formation have occurred in the nuclear region. The first burst occurred about 300\,Myr ago, reaching as close as 0\farcs5 (35\,pc) from the nucleus to the south-west, and  north-east and north-west of the nucleus. The location of this burst may extend to even closer to the nucleus, but the dominance of the BB component in this region may preclude the detection of this component there. Particularly conspicuous is a region between 0\farcs8 (60\,pc) and 1\farcs6 (100\,pc) from the nucleus to the south-west, where it is the only stellar population component. The second burst occurred about 30\,Myr ago, mainly in an approximately ring-like structure with an average radius of $\approx$\,100\,pc.

\begin{figure}
\centering
\includegraphics[scale=0.6]{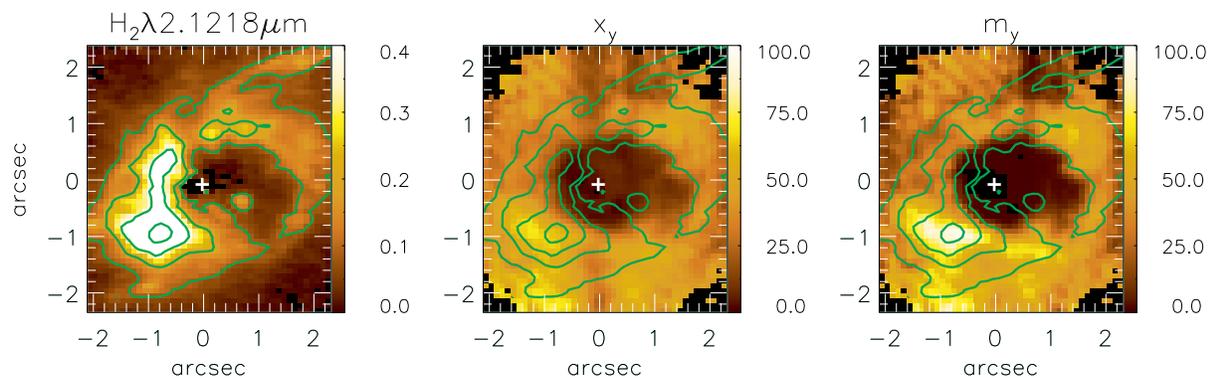}
\caption{Comparison of the H$_2$ flux distribution (left) from \citet{vale12} with the distribution of the 30\,Myr stellar population percent contribution to the flux at 2.1\,$\mu$m (central panel) and percent  contribution by mass (right panel). The green contours on the central and right panels are from the H$_2$ flux distribution.}
\label{comparison}
\end{figure}

In \citet{vale12} we have obtained emission line flux distributions  from the same data we are using here. The flux in the molecular hydrogen emission line H$_2\lambda2.1218\mu$m, shown in the left panel of Fig.\,\ref{comparison} is distributed in a ring -- first shown by \citet{mulsan09} -- similarly to what is observed in NGC\,4151 \citep{sb10}. In Fig.\,\ref{comparison}, the H$_2$ ring (left panel) can be compared with the young stellar population distributions $x_y$ and $m_y$ (central and right panels, respectively). This figure shows that $x_y$ and $m_y$ are enhanced at locations  covered by the H$_2$ ring, and, in particular, the region of strongest H$_2$ emission is the region with the largest contribution of the young stellar component.

\begin{figure}
\centering
\includegraphics[scale=0.6]{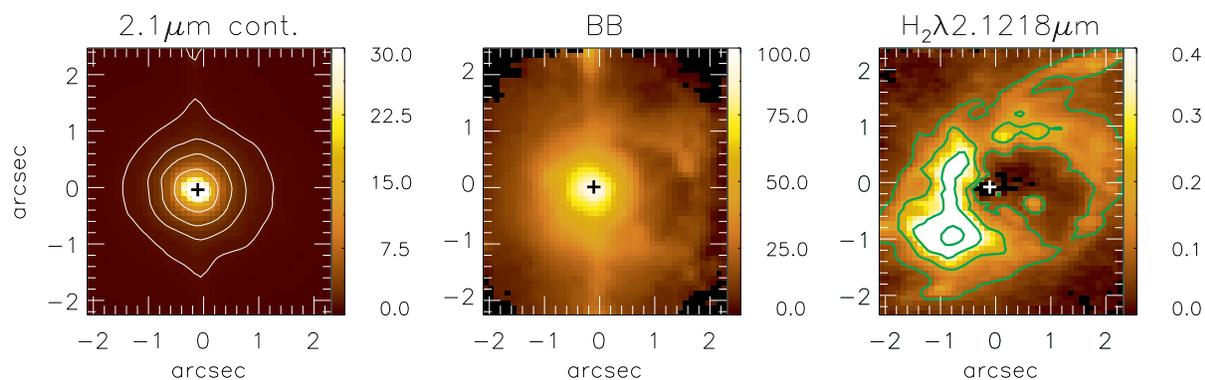}
\caption{Comparison of  the continuum flux distribution (left panel) with the distribution of the percent contribution of the BB components (central panel) and the flux distribution of H$_2$  (right panel) from \citet{vale12}.}
\label{comparison1}
\end{figure}

The H$_2$ ring can be identified with the circumnuclear disk of molecular gas first reported by \citet{schinnerer00}. In Barbosa et al. (in preparation), we show that the H$_2$ emitting gas in this ring seems to be expanding in the plane of the galaxy, confirming  previous suggestions by \citet{galliano02} and  results from observations with the instrument SINFONI on the VLT \citep{mulsan09}. Recent observations of the molecular gas emission  in the mm spectral region (e.g. lines of CO, HCN) by \citet{krips11} also show this ring as well as its expansion, which, combined with the measured line ratios, have led these authors to propose a scenario in which part of the molecular gas is radially blown outwards as a result of shocks. \citet{krips11} suggest that the shocks originate in the interaction of the AGN jet with the molecular gas.

\citet{krips11} also point out that an expanding shell of shocked gas can originate in stellar winds from the evolution of a super stellar cluster (although they do not favor this interpretation). This is the hypothesis put forth by \citet{sakamoto06} to explain the presence of ``superbubbles" observed in molecular gas emission in starburst galaxies such as NGC\,253 and M\,82 \citep{matsushita00}.  In the case of AGN, the presence of superbubbles  has been reported by Lipari and collaborators in a number of galaxies, examples being Mrk\,231 \citep[e.g.][and references therein]{lipari05} and IRAS\,04505-2958\citep{lipari09}. These authors suggest that broad absorption line (BAL) systems seen in the spectra of AGN could be due to these superbubbles -- giant expanding gas shells which originate in circumnuclear starbursts. \citet{lipari06} propose an  evolutionary scenario in which massive circumnuclear starbursts in the vicinity of AGNs produce supernovae and hypernovae which lead to these large-scale expanding shells, which are often obscured by dust. 
Later in the evolution, as the starburst ages and the shells expand,  the AGN becomes the main source of ionization of the circumnuclear gas.
This scenario is shown to reproduce many of the observed correlations among spectral properties of infrared-luminous AGNs. For lower luminosity Seyfert galaxies, \citet{sb01} also propose such an evolutionary scenario to explain the excess of young and intermediate age stellar population within the inner kiloparsec: the gas which flows in, first triggers star formation in the circumnuclear region, and the nuclear activity is then subsequently triggered.

Thus, one possibility to be considered is that the H$_2$ ring is actually a superbubble, for which the strongest emission comes from the galaxy plane, where there is more gas to be swept by the bubble. This is supported by the fact that  NGC\,1068 sits next to Mrk\,231 in the diagram of \citet{davies07} which relates the AGN luminosity with the circumnuclear starburst age. Although the expansion velocity of the H$_2$ ring -- under 100\,\kms (Barbosa et al., in preparation) -- is  much lower than in Mrk\,321 superbubbles, in nearby starbursts, observations  in the milimetric spectral region show similar expansion velocities of a few tens of km\,s$^{-1}$.  The morphology and sizes (100--200\,pc) of superbubbles in nearby starbursts are also similar to those of the molecular ring of NGC\,1068.  One difference is the mass: while in the nearby starbursts the gas masses are of the order of 10$^6$ solar masses (e.g. in NGC\,253), the molecular gas mass in the inner $\approx$\,100\,pc of NGC\,1068 is estimated to be an order of magnitude larger \citep[][and references therein]{mulsan09}.

In the case of the nearby starburst galaxy  NGC\,253,  \citet{sakamoto06} have calculated the necessary power to drive the outflow and concluded that a 10$^6$\,M$_\odot$ super star cluster could be the origin of such bubbles, created by successive supernova explosions, which can have a duration of $\approx$\,30\,Myr, arguing that these bubbles are the end products of star formation. Having a mass an order of magnitude larger than the super-bubbles, it is not clear that the H$_2$ ring in NGC\,1068 has a similar origin. 

It is interesting to note that the apparent center of the H$_2$ ring does not coincide with the nucleus, but is located  in the region $\approx$\,1\arcsec\ to the south-west where the 300\,Myr stellar population is particularly enhanced. If the H$_2$ ring were actually a superbubble originated from stellar winds and supernova explosions,  a cluster located there -- in the center of the ring -- could have originated it. A problem with this scenario is the fact that the ring is expanding and, even at a lower limit of the expansion velocity of 10\,km\,s$^{-1}$, it would reach its present radius of about 100\,pc in only 10\,Myr. 
As supernovae appear no later than $\approx$\,50\,Myr,  a cluster of age  300\,Myr seems too old to have originated the H$_2$ ring in NGC\,1068.

And why would  $x_y$ be enhanced in the H$_2$ ring? Why are the regions with strongest H$_2$ emission (e.g. at $\approx$\,1\farcs5 east of the nucleus) the ones with highest concentration of young stars? Could an expanding bubble have triggered the 30\,Myr episode of star formation? This seems unlikely due to the fact pointed out above that in 10\,Myr the ring moves at least $\approx$\,100\,pc, and this is not consistent with the present radius of the ring, of $\approx$\,100\,pc and the age of the stars in the ring, of 30\,Myr. 

On the other hand, the 30\,Myr age is consistent with the presence of supernova explosions originated in this same population, which could be the origin of the shocks inferred by \citet{krips11} from the molecular line ratios. The presence of such supernova explosions could also explain the high velocity dispersion observed in the region of strongest H$_2$ emission at 1\farcs5  east of the nucleus (Barbosa et al. in preparation), and could be the origin of the apparent expansion of the H$_2$ ring. A spatial coincidence between a 10--50\,Myr stellar population and strong H$_2$ emission was also found by \citet{martins10} at 100\,pc south of the nucleus in a recent long-slit study of the stellar population of NGC\,1068.

\subsubsection{Older age components and continua}

A comparison between the stellar velocity dispersion ($\sigma_\star$) distribution (Fig.\ref{stel}) and the distribution of $x_{yi}$ and $x_{io}$ in Fig.\,\ref{pop} reveals that, in most of the region where there is a drop in $\sigma_\star$, the dominant stellar population is the intermediate age one ($yi$ and $io$). We tentatively conclude that this intermediate age stellar population is related to the $\sigma_\star$ drop, as we have found already for Mrk\,1066 \citep{riffel10} and Mrk\,1157 \citep{riffel11}. Similar rings seem to be common in the central region of active galaxies \citep[e.g.][]{barbosa06,riffel08,n7582} and can be  interpreted as being due to colder regions with more recent star formation than the underlying bulge. This interpretation is in agreement with the previous analysis of SINFONI data by \citet{davies07}, who have estimated as 200--300\,Myr the age of the stars in the nuclear compact disk of NGC\,1068. In a previous work, \citet{thatte97a}  have claimed to have found a stellar core with intrinsic size scale of 45\,pc. Using the estimated mass-to-light ratio and assuming some contamination from old stars, they propose an age of 500\,Myr for a young age component in this region, in agreement with our results.

We have found a small contribution from the old component, $x_o$, to the flux in the inner $\approx$\,0\farcs5 (35\,pc); but even though the flux contribution in the near-IR is not larger than $\approx$\,25\%, the mass contribution is close to $\approx$\,100\%. The presence of an old stellar component around the nucleus, with age\,$\ge\,10^9$yr, has already been suggested by \citet{crenkra00}, on the basis of long-slit STIS spectra, as the dominant light component in the optical part of the spectrum.

Fig.\,\ref{pop-ebv} shows the distribution of the power-law and black-body components. The power-law component does not contribute much at the nucleus, being spread around it. The vertical line is an artifact, which appears also in the lowest temperature black-body component; it is an instrumental effect, due to scattered nuclear light along the vertical mirror slices of NIFS. We attribute the extra-nuclear contribution of the power-law component  to scattered light from the nucleus, which has been found to be a strong component of the continuum in the ultraviolet \citep{crenkra00}.

As pointed out above, the black-body components dominate completely the light at the nucleus, in particular the ones with temperatures in the range 700--800\,K (Fig.\,\ref{pop-ebv}), comprising almost 100\% of the flux. We interpret them as originating in the dusty torus surrounding the AGN, which is unresolved by our observations \citep{marco00,jaffe04,raban09,martins10}. In Fig.\,\ref{comparison1}, we compare the continuum emission in the K band (left) with the sum of all black-body components (center) and with the H$_2$ flux distribution (right). This comparison -- which shows that the continuum emission and BB flux distributions are similarly concentrated --  gives support to the identification of the peak of the K-band continuum with the black-body components contribution. Some contribution of the black-body components is seen also beyond the inner arcsecond, which we attribute to the presence of faint dust emission or to further contribution from scattered nuclear light spread in the nuclear region. The comparison between the central and right panels of Fig.\,\ref{comparison1} shows that some BB emission is seen also along the west part of the ring (to the right in Fig.\,\ref{comparison1}).

\section{Conclusions}

We have mapped the stellar kinematics and age components, as well as the contribution of power-law and black-body emission components in the inner 180\,pc radius of NGC\,1068 at a spatial resolution of 8\,pc using integral field spectroscopy in the near-infrared. The main results of this paper are:

\begin{itemize}

\item The stellar kinematics are well described by rotation with a projected amplitude of $\approx$\,45\,km\,s$^{-1}$ and major axis a position angle of $\approx$\,80$\deg$;

\item The stellar velocity dispersion $\sigma_\star$ shows a drop from 165\,km\,s$^{-1}$ to 125\,km\,s$^{-1}$ within the inner 1\arcsec -- 70\,pc at the galaxy;

\item The dominant stellar populations over the inner 180\,pc radius are the young (mainly 30\,Myr) and intermediate age (mainly 300\,Myr) ones;

\item Within the inner 0\farcs5 (35\,pc), the black-body components with 700$\le$\,T$\le $800\,K dominate the flux, and are attributed to the dusty torus (unresolved by our observations) surrounding the active nucleus;

\item Within the inner 35\,pc, the old component (age$\ge$\,5\,Gyr) dominates the contribution of the stellar population in terms of mass fraction;

\item The distribution of the youngest (30\,Myr) stellar population is co-spatial with an expanding ring of H$_2$ emission at $\approx$\,100\,pc from the nucleus. We speculate that supernova explosions from this young stellar population could be the origin of the apparent expansion of the ring;


\item In the region of the $\sigma_\star$-drop, the dominant stellar population is the intermediate age (mainly 300\,Myr) one.

\end{itemize}

NGC\,1068 -- the best studied Seyfert\,2 galaxy is thus another case of an active nucleus surrounded by recent star formation in a circumnuclear ring, supporting the evolutionary scenario for AGN \citep{sb01,lipari06,davies07}. In such a scenario, gas from the external regions of the galaxy (or even acquired from outside)  flows to the nuclear region, first triggering star formation in the inner few hundred parsecs, and subsequently reaches the nucleus to feed the supermassive black hole, triggering the nuclear activity. 


\acknowledgments We thank the anonymous referee for valuable suggestions which helped to improve the paper. Based on observations obtained at the Gemini Observatory, which is operated by the Association of Universities for Research in Astronomy, Inc., under a cooperative agreement
with the NSF on behalf of the Gemini partnership: the National Science Foundation (United
States), the Science and Technology Facilities Council (United Kingdom), the
National Research Council (Canada), CONICYT (Chile), the Australian Research Council
(Australia), Minist\'erio da Ci\^encia e Tecnologia (Brazil) 
and Ministerio de Ciencia, Tecnologia e Innovaci\'on   (Argentina). This work has been partially supported by the Brazilian intitution CNPq. RR thanks the Brazilian funding agency FAPERGS (process 11/1758-5) for
financial support.

{}   
\clearpage
\end{document}